\documentclass[useAMS,usenatbib,usegraphicx]{mn2e}

\title[Origin of strong magnetic fields in Milky-Way like galactic haloes]{Origin of strong magnetic fields in Milky-Way like galactic haloes}
\author[A. M. Beck et al.]{A. M. Beck$^{1,2}$\thanks{E-mail: abeck@usm.uni-muenchen.de}, H. Lesch$^{1}$, K. Dolag$^{1,3}$, H. Kotarba$^{1}$, A. Geng$^{4}$ and F. A. Stasyszyn$^{1}$\\
  $^{1}$University Observatory Munich, Scheinerstr. 1, D-81679 Munich, Germany\\
  $^{2}$Max Planck Institute for Extraterrestrial Physics, Giessenbachstr. 1, D-85748 Garching, Germany\\
  $^{3}$Max Planck Institute for Astrophysics, Karl-Schwarzschild-Str. 1, D-85741 Garching, Germany\\
  $^{4}$University of Konstanz, Department of Physics, Universitaetsstr. 10, D-78464 Konstanz, Germany}

\usepackage[latin1]{inputenc}
\usepackage{units}
\usepackage{float}
\usepackage{graphicx}
\usepackage{amsmath} 
\usepackage{amssymb}
\usepackage{amsfonts}
\usepackage{xcolor}
\usepackage{times}
\usepackage{ulem}

\begin{document}

\date{Accepted 2012 February 15. Received 2012 February 09; in original form 2011 November 10}

\pagerange{\pageref{firstpage}--\pageref{lastpage}} \pubyear{2012}

\maketitle

\label{firstpage}


\begin{abstract}
\noindent{}An analytical model predicting the growth rates, the absolute growth times and the saturation values of the magnetic field strength within galactic haloes is presented.
The analytical results are compared to cosmological MHD simulations of Milky-Way like galactic halo formation performed with the N-body / \textsc{Spmhd} code \textsc{Gadget}.
The halo has a mass of $\approx{}3\cdot{}10^{12}$ $M_{\odot}$ and a virial radius of $\approx{}$270 kpc.
The simulations in a $\Lambda$CDM cosmology also include radiative cooling, star formation, supernova feedback and the description of non-ideal MHD.

\noindent{}A primordial magnetic seed field ranging from $10^{-10}$ to $10^{-34}$ G in strength agglomerates together with the gas within filaments and protohaloes.
There, it is amplified within a couple of hundred million years up to equipartition with the corresponding turbulent energy.
The magnetic field strength increases by turbulent small-scale dynamo action.
The turbulence is generated by the gravitational collapse and by supernova feedback.
Subsequently, a series of halo mergers leads to shock waves and amplification processes magnetizing the surrounding gas within a few billion years.
At first, the magnetic energy grows on small scales and then self-organizes to larger scales.
Magnetic field strengths of $\approx{}10^{-6}$ G are reached in the center of the halo and drop to $\approx{}10^{-9}$ G in the IGM.

\noindent{}Analyzing the saturation levels and growth rates, the model is able to describe the process of magnetic amplification notably well and confirms the results of the simulations.
\end{abstract}


\begin{keywords}
cosmology: early universe -- galaxies: formation -- galaxies: halos -- galaxies: magnetic fields -- methods: analytical -- methods: numerical 
\end{keywords}


\section{Introduction}

\noindent{}The Lambda Cold Dark Matter model ($\Lambda$CDM) is the standard tool describing the evolution of the universe \citep{komatsu11}.
Quantum fluctuations in the primordial energy distribution develop into the condensing matter and trigger gravitational instabilities.
Dark matter clumps in filaments and protohaloes, and subsequently baryonic matter falls into the potential wells of the dark matter, thereby forming the first stars and galaxies.
In a hierarchical process of merger events, larger structures grow \citep{white78,white91}.
Numerical simulations of structure formation within a $\Lambda$CDM universe show good agreement between the calculated and the observed distribution of matter and structures \citep{springel05b, springel06}.
However, cosmic magnetic fields are still widely neglected in these kind of simulations, although their presence can influence the dynamics of an astrophysical system significantly.

\noindent{}Observations reveal strong magnetic fields of $\mu$G strength in late-type galaxies
(for reviews on cosmic magnetism see e.g. \cite{beck96}, \cite{widrow02}, \cite{kulsrud08}, and references therein).
The energy density of these magnetic fields is comparable to other dynamically important energy densities, i.e. the magnetic field seems to be in equipartition with them.
\cite{neronov10} also find strong magnetic fields permeating the intergalactic medium (IGM).
The IGM magnetic fields are highly turbulent \citep{ryu08} and their strength is estimated to the order of nG \citep{kronberg08}.
Additionally, magnetic fields of $\mu$G strength can be found in high redshift galaxies \citep{bernet08}.
Also, there is evidence of highly magnetized damped Lyman alpha systems at redshift $\approx{}2$, which act as building blocks for galactic systems \citep{wolfe05}.

\noindent{}The origin of these magnetic fields is still unclear.
Global primordial magnetic fields can be seeded by battery processes in the early universe \citep{biermann50,mishustin72,zeldovich83,huba93}.
Alternatively, seed fields can be generated by phase transitions after the BigBang or various other mechanisms \citep[see][for a review]{widrow02}.

\noindent{}In a subsequent process, these weak seed fields of sometimes $\leq{}10^{-20}$ G have to be amplified to the observed values.
\cite{lesch95} demonstrate the possibility of strong magnetic fields at high redshifts through battery processes and protogalactic shear flow amplification.
The presence of strong $\mu$G galactic magnetic fields is commonly explained by galactic dynamos converting angular momentum into magnetic energy in differentially rotating disks.
The two main theories are the $\alpha$-$\omega$ dynamo \citep{ruzmaikin79} or the cosmic ray driven dynamo \citep{lesch03, hanasz09}.
For reviews of dynamo theory see \cite{brandenburg05} or \cite{shukurov07}.
However, these dynamos operate on timescales ($e$-folding time, not absolute amplification time) of the order of $10^{8}$ yrs and require a differentially rotating galactic disk.
Hence, irregular galaxies at high redshift have to be magnetized by another process.
Another possible magnetization process is the ``Cosmic dynamo'' as given by \cite{dubois10}.
Within their approach, the universe is magnetized by gravitational instabilies, galactic dynamos and wind-driven outflows of gas at times of violent star formation activity.

\noindent{}Small-scale dynamos operate on timescales of the order of $10^{6}$ yrs through random and turbulent shear flow motions \citep{biermann51}.
Magnetic energy increases exponentially on small scales first by stretching, twisting and folding the magnetic field lines by random motions and then organizing them on the largest turbulent eddy scale \citep{zeldovich83,kulsrud92,kulsrud96,malyshkin02,schekochihin02,schekochihin04,schleicher10}.
Galaxy mergers are a natural part of the bottom-up picture of the growth of structures in the universe.
\cite{kotarba10,kotarba11} and \cite{geng12} show that turbulence induced during galactic major and minor mergers is able to amplify magnetic fields in galaxies and in the IGM up to equipartition between the magnetic and turbulent energy density, as expected from the small-scale dynamo theory.
This theory is a good method to describe the amplification processes and corresponding timescales \citep[e.g.][]{arshakian09}.
However, galactic dynamos are still inevitable to explain the regularity of galactic magnetic fields and their spiral structure, which are revealed by observations.

\noindent{}Analytical calculations and cosmological simulations of structure formation including the evolution of magnetic fields can give new insights in the physical processes of creating, amplifying and saturating magnetic fields in the universe on all kind of scales.

\noindent{}In this work, an analytical model predicting the growth rates, the absolute growth times and the saturation values of the magnetic field strength within galactic haloes is presented.
The analytical results are compared to cosmological MHD simulations of Milky-Way like galactic halo formation including star formation and non-ideal MHD.
It is shown that the analytical model and the cosmological simulations agree notably well for different initial, primordial magnetic seed fields spanning a range of 25 orders of magnitudes.

\noindent{}The paper is organized as follows:
The analytical calculations are shown in section 2.
Section 3 briefly describes the numerical method.
In section 4 the cosmological inital conditions and the magnetic seed field are motivated.
A detailed analysis of the performed simulations and the magnetic field amplification is given in section 5.
Section 6 compares the numerical results with the analytical description.
The main results are summarized in section 7.


\section{Analytical description}\label{sec:analytical}

\noindent{}This section gives an analytical approach describing the behaviour of the magnetic field strength during halo formation.
In order to derive an analytical model, expressions for the cosmological decay, the exponential amplification process, the saturation and the relaxing decay of the magnetic field strength are needed.
For large hydrodynamical Reynolds numbers, a stationary flow transits from the laminar regime into the turbulent regime and becomes unstable.
Hence, an overview of the characteristics of a non-stationary perturbated magnetic field in such an unstable flow in a cosmological context is given.


\subsection{Local perturbation ansatz}

\noindent{}The Reynolds number is a characteristic dimensionless quantity describing the ratio of inertial forces and viscous forces of a flow:

\begin{equation}Re=\frac{\textnormal{inertial forces}}{\textnormal{viscous forces}}=\frac{VL}{\nu},\label{equ:reynolds}\end{equation}

\noindent{}with $V$ and $L$ being typical velocity and length scales of the flow and $\nu$ the kinematic viscosity.
A stationary flow will become unstable if the Reynolds number $Re$ exceeds a critical value $Re_\rmn{crit}$.
In this case, any initial infinitesimal perturbation will be amplified through the flow.
In the following calculations, the hydrodynamical flow is assumed to be unstable, since the inertial length scale is sufficiently larger than the viscous length scale.
The magnetic Reynolds number $Rm$, which is defined as $Rm=VL/\eta$ (with the magnetic resistivity $\eta$), is also sufficiently large to allow for perturbations to grow.
The induction equation of ideal MHD reads:

\begin{equation}\frac{\partial{}\bmath{B}(\bmath{x},t)}{\partial{}t}=\bmath{\nabla}\times\left[\bmath{v}(\bmath{x},t)\times\bmath{B}(\bmath{x},t)\right].\label{equ:induction}\end{equation}

\noindent{}Within the small perturbation approximation, the velocity field $\bmath{v}(\bmath{x},t)$ and the magnetic field $\bmath{B}(\bmath{x},t)$, respectively, can be decomposed into the sum of a stationary component $\bmath{v}_{0}(\bmath{x},t)$ and $\bmath{B}_{0}(\bmath{x},t)$, and a perturbated component $\bmath{v}_{1}(\bmath{x},t)$ and $\bmath{B}_{1}(\bmath{x},t)$, respectively.
For the induction equation (\ref{equ:induction}), this procedure results in the following decomposition:

\begin{equation}\frac{\partial{}\bmath{B}_{0}}{\partial{}t}=\bmath{\nabla}\times\left[\bmath{v}_{0}\times\bmath{B}_{0}\right],\end{equation}

\begin{equation}\frac{\partial{}\bmath{B}_{1}}{\partial{}t}=\bmath{\nabla}\times\left[\bmath{v}_{0}\times\bmath{B}_{1}+\bmath{v}_{1}\times\bmath{B}_{0}\right].\label{equ:ind_pert}\end{equation}

\noindent{}The right crossproduct in Eq. (\ref{equ:ind_pert}) can be dropped, because only the growth of magnetic perturbations in a stationary flow is important, and the growth of velocity perturbations in a weak stationary magnetic field can be neglected.
The general solution of Eq. (\ref{equ:ind_pert}) is a sum of special solutions, whereby $\bmath{B}_{1}$ includes a time-dependent factor $e^{-i\Omega{}t}$.
The complex frequency $\Omega$ is given by $\Omega=\omega+i\Gamma$, with periodicity $\omega$ and growth rate $\Gamma$.
For growing perturbations $\Gamma$ will be positive.
The perturbated component $\bmath{B}_{1}$ can be further decomposed into

\begin{equation}\bmath{B}_{1}(\bmath{x},t)=B_{t}(t)\bmath{s}(\bmath{x}),\end{equation}

\begin{equation}B_{t}(t)=B_{t}(t_{0})e^{\Gamma{}t}e^{-i\omega{}t},\end{equation}

\noindent{}with a spatially dependent complex vector function $\bmath{s}(\bmath{x})$ and a complex scalar amplitude $B_{t}(t)$.
The time derivative of the square of the amplitude is:

\begin{equation}\frac{\partial{}B_{t}^{2}(t)}{\partial{}t}=2\Gamma{}B_{t}^{2}(t).\label{equ:growth_pre}\end{equation}

\noindent{}Eq. (\ref{equ:growth_pre}) was also derived for a kinematic dynamo by \cite{kulsrud92} and for a turbulent magnetized dynamo by \cite{malyshkin02} using spectral analysis of the growth of the magnetic energy.
In the weak-field approximation, the flow is able to amplify the frozen-in magnetic perturbations, since the magnetic field is frozen into the velocity field.
Thereby, any information about the original orientation of the magnetic seed field is lost.
As soon as equipartition is reached, the back-reaction of the magnetic field on the velocity field has to be considered and the weak-field approximation breaks.
Then, one can either solve the full Navier-Stokes equations including magnetic pressure and tension forces or model the back-reaction by truncating the growth rate $\Gamma$.
\cite{belyanin93} describe the latter ``Equipartition dynamo'' approach by expanding $\Gamma$ in a power series and considering second-order terms to truncate the growth rate.
These non-linear effects force the maximum amplitude to an equipartition value and the dynamo process saturates.
Generally, a ``Supraequipartition dynamo'' should be expected, because compression of $B^{2} \sim \rho^{\gamma_{\rmn{adi}}}$ (e.g. the effective adiabatic index is $\gamma_{\rmn{adi}}=4/3$ for isotropic compression) can still occur.
This would result in the magnetic energy density first to shoot over the turbulent energy density ($\gamma_{\rmn{adi}}=1$), and then decay towards the saturation level.
However, this effect is too small to be significant.
Now, to account for the saturation of the magnetic field at the equipartition level, $\Gamma$ is amended \citep[][]{belyanin93} as follows

\begin{equation}\Gamma=\gamma\left[1-\frac{B^{2}_{t}(t)}{B^{2}_{\rmn{sat}}}\right],\label{equ:gamma_expand}\end{equation}

\noindent{}and the equation for the growth of the magnetic field amplitude takes the form:

\begin{equation}\frac{\partial{}B_{t}^{2}(t)}{\partial{}t}=2\gamma{}\left[B_{t}^{2}(t)-\frac{B_{t}^{4}(t)}{B^{2}_{\rmn{sat}}}\right].\label{equ:growth_final}\end{equation}

\noindent{}This growth of the magnetic perturbations is an iterative process: $\bmath{B}_{1}$ can only grow to the order of magnitude of $\bmath{B}_{0}$, then a new $\bmath{B}_{0}$ and $\bmath{B}_{1}$ have to be defined.
Furthermore, since turbulence is the driver of the magnetic field amplification, equipartition between the magnetic and the turbulent energy density is a good approximation for the saturation magnetic field (no further growth will occur, once the magnetic field has saturated):

\begin{equation} \frac{B^{2}_{\rmn{sat}}}{8\pi}=\frac{1}{2}\rho{}v^{2}_{\rmn{turb}},\end{equation}

\noindent{}with the density $\rho$ and the turbulent velocity $v_{\rmn{turb}}$ of the system.
Together with the initial condition $B^{2}_{t}(t=0)=B^{2}_{0}$ Eq. (\ref{equ:growth_final}) has the solution \citep[][]{landau59}:

\begin{equation}B_{t}(t)=\frac{1}{\sqrt{(4\pi\rho{}v^{2}_{\rmn{turb}})^{-1}+B^{-2}_{0}e^{-2\gamma{}t}}}.\label{equ:solution_pre}\end{equation}

\noindent{}As a next step the growth rate $\gamma$ has to be determined.
The kinetic energy of the turbulence can be assumed to follow a one-dimensional Kolmogorov spectrum of the form
\begin{equation} I(k)\sim{}v_{\rmn{turb}}^{2}k^{-5/3},\end{equation}
where $v^{2}_{\rmn{turb}}$ is the mean square turbulent velocity.
From Eq. (\ref{equ:growth_pre}) \cite{kulsrud96} find for the growth rate $\gamma$
\begin{equation} 2\gamma\approx\int{\frac{k^{2}I(k)}{kv_{k}}dk}\approx\int{\sqrt{kI(k)}dk},\label{equ:spectrum}\end{equation}
with an eddy turnover rate $kv_{k}\approx{}\left[kI(k)\right]^{1/2}$ and with $v_{k}$ being the typical eddy velocity on $k$ scale.
Integrating Eq. (\ref{equ:spectrum}) yields the growth rate $\gamma$.
Performing additional calculations including resistivity, \cite{kulsrud96} find an advanced version of $\gamma$:
\begin{equation}\gamma=2.050\frac{v^{3/2}_{\rmn{turb}}k^{1/2}_{\rmn{turb}}}{\eta_{\rmn{turb}}^{1/2}}.\label{equ:kulsrud_gamma}\end{equation}

\noindent{}Eq. (\ref{equ:kulsrud_gamma}) describes the turnover rate of the smallest turbulent eddy.
Magnetic resistivity, which is essential for topological changes of the magnetic field lines through reconnection on small scales and transfer of magnetic energy into internal energy, and which is effectively lowering the growth rate, is already included via $\eta_{\rmn{turb}}$.
The dynamo process increases the magnetic energy on the dissipation scale first with a rate of $2\gamma$ and saturates at a value comparable to the turbulent energy on small scales first.
The magnetic energy is then transferred to larger scales with a rate of $\frac{3}{4}\gamma$ \citep{kulsrud96} by Lorentz forces unwrapping the folded field lines on small scales in an inverse cascade process.
Finally, the magnetic field reaches saturation at a value comparable to the turbulent energy in the largest eddy \citep[][]{kulsrud92,malyshkin02}.
The growth rate given by Eq. (\ref{equ:kulsrud_gamma}) is taken from the kinematic dynamo theory \citep[][]{kulsrud96}, but \cite{malyshkin02} find that the calculation leading to Eq. (\ref{equ:kulsrud_gamma}) still holds for the magnetized turbulent dynamo in the weak-field approximation.

\noindent{}These small-scale dynamos operate whenever turbulent and random motions and shear flows are stretching, twisting and folding the magnetic fields lines \citep{zeldovich83,schleicher10}.
Field lines come close on small scales first and hence, the magnetic energy increases at first on small scales.
The amplification is dominated by the turbulent dynamo action and effects resulting from compression can be neglected within this model.


\subsection{Cosmological evolution and decay}

\begin{figure*}
\begin{center}
  \includegraphics[angle=90,width=0.8\textwidth]{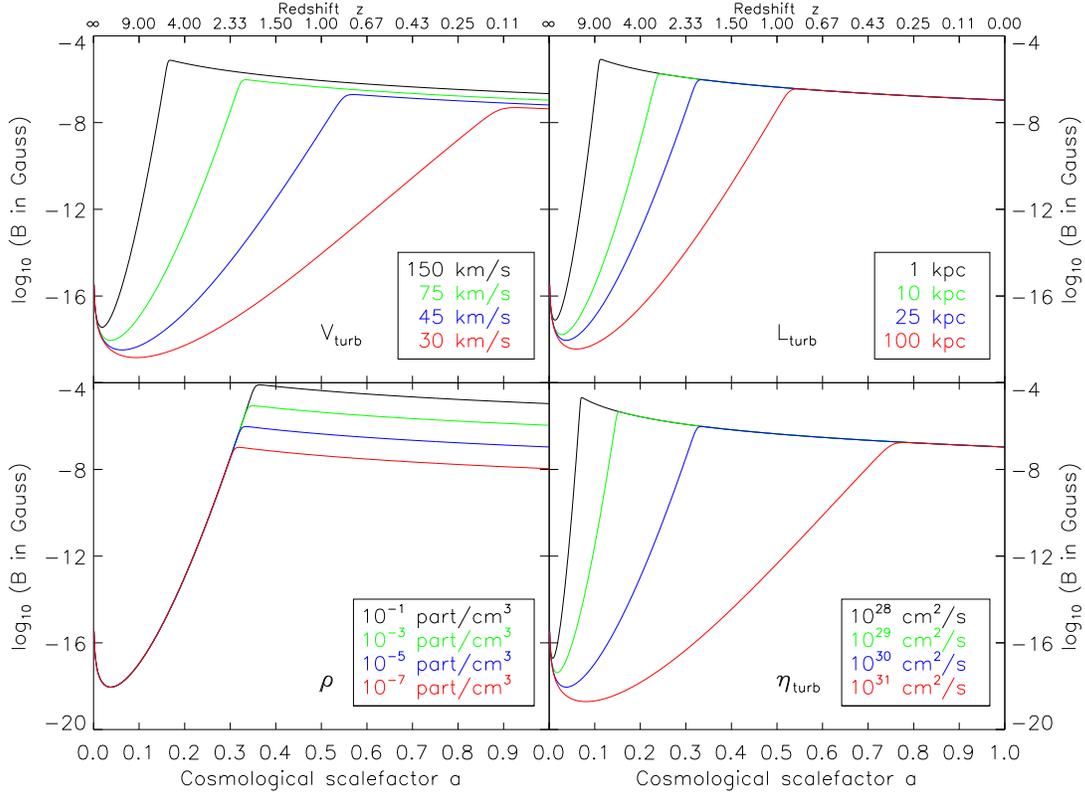}
  \caption{Analytical growth curves for the magnetic field amplitude as a function of redshift for different numerical parameters.
In each panel one parameter is varied, while the other parameters are held constant.
Differences in the growth rates and saturation levels can be seen.}
  \label{fig:model_param}
\end{center}
\end{figure*}

\noindent{}The next step towards an analytical model of the magnetic field amplification is to modify the growth equation (\ref{equ:growth_final}) such that the scalefactor $a(t)$ describing the expansion of the universe during its evolution is accounted for.
For an isotropic and stationary magnetic field, the magnetic flux has to be conserved when space is expanding.
From the first law of thermodynamics, the proportionality between the scalefactor $a$ and the energy density of electromagnetic fields (or ultrarelativistic particles) $\varepsilon{} \sim B_{t}^{2}$, can be derived \citep{longair98}:

\begin{equation}\varepsilon{} \sim a^{-4}.\label{equ:cosmo_b}\end{equation}

\noindent{}Any cosmological equation for the evolution of the magnetic field amplitude has thus to account for the proportionality $B_{t} \sim a^{-2}$.
In cosmological simulations, the scalefacor $a(t)$ instead of physical time $t$ is the natural integration variable.
Given a flat universe without any radiation pressure, the evolution equation for the scalefactor takes the form \citep[e.g.][]{longair98}

\begin{equation}dtH_{0}=da(\frac{\Omega_{M}}{a}+a^{2}\Omega_{\Lambda})^{-\frac{1}{2}},\label{equ:evolv_scale}\end{equation}

\noindent{}with $\Omega_{\rmn{M}}$ and $\Omega_{\Lambda}$ being the density contributions of matter and the cosmological constant, and $H_{\rmn{0}}$ the present-day Hubble constant.
The employed parameters of this $\Lambda$CDM cosmology (Table \ref{tab:cosmo_parameter}) are close to the observed values \citep[see][]{komatsu11}.
With the initial condition $a(0)=0$ Eq. (\ref{equ:evolv_scale}) has the solution:

\begin{equation}t(a)=\frac{2}{3H_{0}\sqrt{\Omega_{\Lambda}}}\cdot\mathrm{asinh}\left(\sqrt{\frac{\Omega_{\Lambda}}{\Omega_{M}}}a^{\frac{3}{2}}\right).\label{equ:solv_scale}\end{equation}

\noindent{}Combining equations (\ref{equ:solution_pre}), (\ref{equ:kulsrud_gamma}), (\ref{equ:cosmo_b}) and (\ref{equ:solv_scale}) finally results in:

\begin{equation}B_{t}(a)=\frac{1}{a^2}\left[(4\pi\rho{}v^{2}_{\rmn{turb}})^{-1}+B^{-2}_{0}e^{-2\gamma{}t(a)}\right]^{-\frac{1}{2}}.\label{equ:solv_final}\end{equation}

\begin{table}
\begin{center}
\renewcommand{\arraystretch}{1.2}
  \begin{tabular}{@{}lll}
    \hline\hline
    &&\hspace*{-3.0cm}\textsc{$\Lambda$CDM Cosmology Parameters}\\\hline\hline
    Matter density & $\Omega_{\rmn{M}}$ & 0.3\\
    Dark energy density & $\Omega_{\rmn{\Lambda}}$ & 0.7\\
    Total density & $\Omega_{\rmn{0}}$ & 1.0\\
    Hubble constant & $H_{0}$ & 70 km s$^{-1}$Mpc$^{-1}$\\\hline\hline
  \end{tabular}
  \caption{Parameters used for the $\Lambda$CDM cosmology.}
  \label{tab:cosmo_parameter}
\end{center}
\end{table}

\noindent{}Eq. (\ref{equ:solv_final}) is an approximation for the growth of the magnetic field strength during the matter-dominated and later phases of the Universe.

\noindent{}In the beginning, the term $a(t)^{-2}$ dominates and results in the cosmological dip at high redshifts.
Then, the magnetic energy increases with $e^{2\gamma{}t(a)}$ during the gravitational collapse and due to star formation induced turbulence.
This growth stops when equipartition is reached, since non-linear effects are truncating $\gamma$.
Finally, when the system is relaxing and the turbulence is decaying, also the magnetic field will decay as $a(t)^{-2}$ corresponding to a power-law decay with $B_{t} \sim t^{-4/3}$.
This decay is slightly stronger compared to the $B_{t} \sim t^{-5/4}$ decay given by \cite{landau59} or \cite{george92} for the final stages of decaying kinematic turbulence (assuming that the magnetic energy decays with the same power-law as the turbulent energy, which is an approximation).
The final stage of decay is reached by the time, when the Reynolds number becomes sufficiently small.
This happens when the back-reaction of the magnetic field on the velocity field suppresses turbulent motions and also magnetic amplification and the system relaxes.
\cite{subramanian06} found the power-law decay to set in already for Reynolds numbers still as large as $Re\approx{}100$ in galaxy clusters.


\subsection{Numerical parameters}

\begin{table}
\begin{center}
\renewcommand{\arraystretch}{1.2}
  \begin{tabular}{@{}lll}
    \hline\hline
    &&\hspace*{-3.0cm}\textsc{Analytical Model Parameters}\\\hline\hline
    Turbulent length & $l_{\rmn{turb}}$ & 25 kpc\\
    Turbulent velocity & $v_{\rmn{turb}}$ & 75 km s$^{-1}$\\
    Turbulent dissipation & $\eta_{\rmn{turb}}$ & $10^{30}$ cm$^{2}$ s$^{-1}$\\
    Gas density & $n_{\rmn{gas}}$ & 10$^{-5}$ cm$^{-3}$\\\hline\hline
  \end{tabular}
  \caption{Parameters used for the calculation of the analytical growth function (physical units).}
  \label{tab:model_parameter}
\end{center}
\end{table}

\noindent{}Eq. (\ref{equ:solv_final}) contains four free parameters: $v_{\rmn{turb}}$, $l_{\rmn{turb}}$, $\eta_{\rmn{turb}}$ and $\rho$.
The numerical values for these parameters have to be taken from observations or extracted from numerical simulations.
Table \ref{tab:model_parameter} shows the values used in this work.
These values correspond to a timescale ($e$-folding time) of $\tau=1/\gamma\approx{}90$ Myr and a saturation value of $B_{\rmn{sat}}\approx{}0.1\mu\rmn{G}$ at redshift zero.
Note, that these numerical parameters are constant mean estimates and do not reflect the time and spatial details of the simulations, but nevertheless are a good approximation.

\noindent{}Estimating values for $v_{\rmn{turb}}$ and $l_{\rmn{turb}}$ is quite challenging, as the densities, velocities and length scales in galactic haloes range over many orders of magnitude.
Therefore, such values can only be associated to typical values within galactic haloes (see section \ref{agreement} for a discussion which values reflect best the simulations).

\noindent{}Fig. \ref{fig:model_param} shows growth curves as a function of redshift for a wide range of values of these parameters.
In each panel, three parameter are held constant (see Table \ref{tab:model_parameter}), while the fourth parameter is varied.
As indicated by the presented calculations, $v_{\rmn{turb}}$, $l_{\rmn{turb}}$ and $\eta_{\rmn{turb}}$ affect the growth rate $\gamma$, while $v_{\rmn{turb}}$ and $\rho$ affect the saturation value $B_{\rmn{sat}}$.
Depending on the parameter configuration, the timescale ranges from order of $10^{6}$ yrs to $10^{8}$ yrs and the saturation value for the magnetic field strength varies from $10^{-5}$ G to $10^{-8}$ G.


\section{Numerical methods}

\noindent{}The simulations in this work are performed with the N-body / \textsc{Spmhd} code \textsc{Gadget} \citep{springel01a,springel05a,dolag09b}.
\textsc{Gadget} uses a formulation of \textsc{Sph}, in which both energy and entropy are conserved \citep{springel02}.
For recent reviews on the \textsc{Sph} and \textsc{Spmhd} methods, see \cite{springel10} and \cite{price12}.
Additionally, \textsc{Subfind} \citep{springel01b,dolag09a} is applied to identify haloes and subhaloes and to calculate their respective center locations and virial radii.

\noindent{}A detailed description of the \textsc{Spmhd} implementation of MHD and its extension to non ideal MHD can be found in \cite{dolag09b} and \cite{bonafede11}.
Here, the standard (direct) \textsc{Spmhd} implementation is used, where the induction equation
\begin{equation}\frac{\partial{}\bmath{B}}{\partial{}t}=\bmath{\nabla}\times\left(\bmath{v}\times\bmath{B}\right)+\eta\Delta\bmath{B}\label{equ:ind}\end{equation}
is evolving the magnetic field, and also a spatially constant magnetic resistivity $\eta$ is applied.
Following \cite{bonafede11}, non-ideal resistivity is assumed to be driven by the turbulence within the gas and $\eta_{\rmn{turb}}$ is of the order of $\approx{}10^{30}$~cm$^{2}$~s$^{-1}$, which is consistent with models of the central regions of galaxy clusters \citep{schlickeiser87,rebusco06}.
Therefore, the constant turbulent resistivity describes the magnetic field decay on sub-resolution scales and is many order of magnitudes larger compared to numerical or ohmic resistivity, which are hence not of interest for this work.

\noindent{}The magnetic field back-reacts on the velocity field via the Lorentz force.
To account for the tensile instability \citep[see][for details]{dolag09b,price12} in \textsc{Spmhd}, the unphysical numerical divergence force is substracted from the equation of motion following an approach by \cite{boerve01}.
Similar to \cite{kotarba10}, a limiter is applied to ensure that the correction force does not exceed the Lorentz force to avoid instabilities.

\noindent{}To ensure a proper evolution of the magnetic field in numerical simulations, it is of fundamental interest to maintain the $\bmath{\nabla}\cdot\bmath{B}$ constraint.
In particular, an erroreous calculation can lead to unphysical sources and sinks of magnetic energy.
The \textsc{Mhd Gadget} code keeps these numerical errors to a minimum.
For a detailed discussion, see \cite{dolag09b} and section \ref{sec:stability}.

\noindent{}The implementation of MHD in \textsc{Gadget} was successfully employed for the study of the magnetic field evolution during star formation \citep{buerzle11a,buerzle11b}, in isolated \citep{kotarba09} and interacting galaxies \citep{kotarba10,kotarba11,geng12}, and in galaxy clusters \citep{donnert09}.

\begin{table}
\begin{center}
  \begin{tabular}{@{}lll}
    \hline\hline
    &\hspace*{-2.8cm}\textsc{Multi-phase Model Parameters}&\\\hline\hline
    Gas consumption timescale & t$_{\rmn{SF}}$ & 2.1 Gyr\\
    Number density threshold & $n_{\rmn{th}}$ & $0.13$ cm$^{-3}$\\
    Mass fraction of massive stars & $\beta$ & $10\%$\\
    Evaporation parameter & A & 1000\\
    Effective supernova temperature & T$_{\rmn{SN}}$ & $10^{8}$ K\\
    Temperature of cold clouds & T$_{\rmn{CC}}$ & 1000 K\\\hline\hline
  \end{tabular}
  \caption{Parameters for the star formation model \citep{springel03a} used in the simulations.}
  \label{tab:sfr_model}
\end{center}
\end{table}

\noindent{}Also, the \cite{springel03a} star formation model is applied.
It describes radiative cooling, UV background heating and supernova feedback in a consistent two-phase sub-resolution model for the interstellar medium.
Cold clouds with a fixed temperature of $T_{\rmn{CC}}$ are embedded into a hot ambient medium at pressure equilibrium.
These cold clouds are evaporating with an efficiency parameter of $A$ and form stars on a timescale of $t_{\rmn{SF}}$, once they reach a density threshold of $\rho_{\rmn{th}}$.
A fraction $\beta$ of these stars is expected to die instantly as supernovae, heating the gas with a temperature of $T_{\rmn{SN}}$.
Additionally, the hot phase is loosing energy via cooling, which is modeled assuming a primordial gas composition (Hydrogen 76\% and Helium 24\%) with a temperature floor of $50$ K \citep[for details see][]{katz96}.
The cooling only depends on density and temperature, but not on metallicity.
This star formation model leads to a self-regulated cycle of cooling, star formation and feedback in the gas.

\noindent{}Table \ref{tab:sfr_model} shows the numerical values of these parameters used in the simulations, which are performed without galactic winds.
These numbers are choosen to reproduce the Kennicutt-Schmidt law between surface density and surface star formation rate \citep{schmidt59,kennicutt98}.

\noindent{}However, for simulations of the turbulent small-scale dynamo the precise details of the star formation scheme are largely unimportant.
Cooling is required to obtain higher gas densities and smaller spatial scales in order to start efficient dynamo action.
Furthermore, feedback-driven turbulence will contribute to the gravitationally-driven turbulence and raise the growth rates of the magnetic field strength.


\section{Setup}

\begin{table*}
\begin{center}
  \begin{tabular}{@{}llllllll}
    \hline\hline
    &&&&\hspace*{-1cm}\textsc{Simulation setup}&\\
    \hline\hline
    Scenario & SF/Cool. & $\rmn{N}_{\rmn{Gas}}$ & $\rmn{N}_{\rmn{DM}}$ & $\rmn{M}_{\rmn{Gas}}$ & $\rmn{M}_{\rmn{DM}}$ & $\rmn{B}^{\rmn{orientation}}_{\rmn{start}}$ & $\rmn{B}^{\rmn{strength}}_{\rmn{start}}$\\
    \hline
    ga0\_bx0 & yes & 68323 & 68323 & $2.6\cdot{}10^{7}$ $\rmn{M}_{\odot}$ & $1.4\cdot{}10^{8}$ $\rmn{M}_{\odot}$ & - & 0 G\\
    ga0\_bx10 & yes & 68323 & 68323 & $2.6\cdot{}10^{7}$ $\rmn{M}_{\odot}$ & $1.4\cdot{}10^{8}$ $\rmn{M}_{\odot}$ & x & $10^{-10}$ G\\
    ga0\_bx14 & yes & 68323 & 68323 & $2.6\cdot{}10^{7}$ $\rmn{M}_{\odot}$ & $1.4\cdot{}10^{8}$ $\rmn{M}_{\odot}$ & x & $10^{-14}$ G\\
    ga0\_bx18 & yes & 68323 & 68323 & $2.6\cdot{}10^{7}$ $\rmn{M}_{\odot}$ & $1.4\cdot{}10^{8}$ $\rmn{M}_{\odot}$ & x & $10^{-18}$ G\\
    ga0\_by18 & yes & 68323 & 68323 & $2.6\cdot{}10^{7}$ $\rmn{M}_{\odot}$ & $1.4\cdot{}10^{8}$ $\rmn{M}_{\odot}$ & y & $10^{-18}$ G\\
    ga0\_bx22 & yes & 68323 & 68323 & $2.6\cdot{}10^{7}$ $\rmn{M}_{\odot}$ & $1.4\cdot{}10^{8}$ $\rmn{M}_{\odot}$ & x & $10^{-22}$ G\\
    ga0\_bx26 & yes & 68323 & 68323 & $2.6\cdot{}10^{7}$ $\rmn{M}_{\odot}$ & $1.4\cdot{}10^{8}$ $\rmn{M}_{\odot}$ & x & $10^{-26}$ G\\
    ga0\_bx30 & yes & 68323 & 68323 & $2.6\cdot{}10^{7}$ $\rmn{M}_{\odot}$ & $1.4\cdot{}10^{8}$ $\rmn{M}_{\odot}$ & x & $10^{-30}$ G\\
    ga0\_bx34 & yes & 68323 & 68323 & $2.6\cdot{}10^{7}$ $\rmn{M}_{\odot}$ & $1.4\cdot{}10^{8}$ $\rmn{M}_{\odot}$ & x & $10^{-34}$ G\\
    \hline
    ga1\_bx0 & yes & 637966 & 637966 & $2.8\cdot{}10^{6}$ $\rmn{M}_{\odot}$ & $1.5\cdot{}10^{7}$ $\rmn{M}_{\odot}$ & - & 0 G\\
    ga1\_bx18 & yes & 637966 & 637966 & $2.8\cdot{}10^{6}$ $\rmn{M}_{\odot}$ & $1.5\cdot{}10^{7}$ $\rmn{M}_{\odot}$ & x & $10^{-18}$ G\\
    ga1\_bx18\_nosf & no & 637966 & 637966 & $2.8\cdot{}10^{6}$ $\rmn{M}_{\odot}$ & $1.5\cdot{}10^{7}$ $\rmn{M}_{\odot}$ & x & $10^{-18}$ G\\
    \hline
    ga2\_bx0 & yes & 5953033 & 5953033 & $3.0\cdot{}10^{5}$ $\rmn{M}_{\odot}$ & $1.6\cdot{}10^{6}$ $\rmn{M}_{\odot}$ & - & 0 G\\
    ga2\_bx18 & yes & 5953033 & 5953033 & $3.0\cdot{}10^{5}$ $\rmn{M}_{\odot}$ & $1.6\cdot{}10^{6}$ $\rmn{M}_{\odot}$ & x & $10^{-18}$ G\\
    \hline\hline
  \end{tabular}
  \caption{Setup of the different simulations: The table lists wether star formation and cooling (SF/Cool.) is applied, the number of gas ($\rmn{N}_{\rmn{Gas}}$) and dark matter particles ($\rmn{N}_{\rmn{DM}}$), the mass of the gas ($\rmn{M}_{\rmn{Gas}}$) and dark matter particles ($\rmn{M}_{\rmn{DM}}$), as well as the initial magnetic field orientation and strength for all simulated scenarios, respectively.}
  \label{tab:setup_sims}
\end{center}
\end{table*}


\subsection{Dark matter initial conditions}

\noindent{}The presented simulations start from cosmological initial conditions introduced by \cite{stoehr02}.
Starting point is a large $\Lambda$CDM dark matter only simulation box run with the \textsc{Gadget} code at different resolutions.
The index of the power spectrum of the initial fluctuations is $n=1$ and the fluctuation amplitude is $\sigma_{8}=0.9$ \citep[see][for details]{stoehr02}.
In a ``typical'' region of the universe a Milky-Way like dark matter halo is identified.
The resulting simulations (with increasing resolution) are labelled GA0, GA1 and GA2, and contain 13~603, 123~775 and 1~055~083 dark matter particles, respectively, inside $R_{200}$, which is the radius enclosing a mean density 200 times the critical density (virial radius).

\noindent{}The forming dark matter halo is comparable to the halo of the Milky-Way in mass ($\approx3\times10^{12}M_{\odot}$) and size ($\approx 270$kpc).
The halo was selected to have no major merger after a redshift of $\approx1$ and also a subhalo population comparable to the satellite population of the Milky-Way was found.
More details about the properties of this halo can be found in \cite{stoehr02}, \cite{stoehr03} and \cite{stoehr06}.
Hence, GA0, GA1 and GA2 provide ideal initial conditions to investigate the evolution of magnetic fields in a galactic halo, similar to the Milky-Way.


\subsection{Gas and magnetic field}

\noindent{}To add a baryonic component, the high resolution dark matter particles are split into an equal amount of gas and DM particles.
The mass of the initial DM particle is splitted according to the cosmic baryon fraction, conserving the center of mass and the momentum of the parent DM particle.
The new particles are displaced by half the mean interparticle distance.

\noindent{}The origin of cosmic magnetic fields is still unclear.
During the early evolution of the universe, magnetic seed fields must have been generated by non-ideal mechanisms, which are independent of the magnetic field itself.
It this work, primordial magnetic fields permeating the entire universe are assumed to be generated by battery processes \citep{biermann50,mishustin72,zeldovich83,huba93} onset of structure formation.
When deriving the induction Eq. (\ref{equ:ind}) of ideal MHD, the fluid is treated as a one particle-type plasma.
Hence, in Ohm's law $\sigma\bmath{E}=\bmath{j}$ (with $\bmath{E}$ the electric field and $\sigma$ the conductivity of the plasma), the current density $\bmath{j}$ is only described by the motion of the protons.
Strictly, the plasma also contains electrons and the current density $\bmath{j}$ is a combination of the proton current density $\bmath{j}_{\rmn{p}}$ and an electron current density $\bmath{j}_{\rmn{e}}$.
Then, electrons and protons are moving at different speeds in the plasma, reacting differently to perturbations.
This results in currents and a non-ideal term in the induction equation of the form:

\begin{equation}\left(\frac{\partial{}\bmath{B}(\bmath{x},t)}{\partial{}t}\right)_{\rmn{seed}}=-c\frac{\bmath{\nabla}n_{\rmn{e}}\times\bmath{\nabla}p_{\rmn{e}}}{n_{\rmn{e}}^{2}e},\label{equ:seeding}\end{equation}

\noindent{}with the speed of light $c$, the elementary charge $e$, the electron density $n_{\rmn{e}}$, and the electron pressure $p_{\rmn{e}}$.
In shocks or non-isotrop regions, the gradients of $n_{\rmn{e}}$ and $p_{\rmn{e}}$ may be non-parallel and a magnetic seed field of the strength $\approx{}10^{-18}$ G may be generated \citep{biermann50,mishustin72,zeldovich83,huba93}.
To seed the magnetic field within the simulations, a magnetic field vector is given to every gas particle pointing into the same direction carrying the same amplitude (most used choice: $10^{-18}$ G in $x$ direction).
This uniform setup gives an initially divergence-free magnetic field inside the simulation volume.

\noindent{}Actually, the magnetic energy should be distributed to the different scales of the simulation, resulting in a magnetic spectrum.
However, this spectral property of the magnetic field can be neglected for magnetic energy densites $\varepsilon_{\rmn{mag}} = B^{2}/8\pi$ sufficiently smaller than the kinetic energy density $\varepsilon_{\rmn{kin}} = \rho{}v^{2}/2$ (weak-field approximation).
In this limit, the magnetic field $\bmath{B}$ will be frozen into the velocity field $\bmath{v}$ and follow its evolution.
Hence, only the amplitude of the magnetic seed field is relevant, but not its direction.

\noindent{}Table \ref{tab:setup_sims} shows the performed simulations, which can be grouped into three sets:
First, low-resolution simulations GA0 are used for a numerical study of different seed field strengths ranging from $10^{-10}$ G to $10^{-34}$ G and being uniform in $x$ direction.
Additionally, a run of GA0 with a seed field in $y$ direction is shown to confirm the neglectability of the magnetic seed field direction.
Second, for the seed field with the strength of $10^{-18}$ G, higher resolution runs GA1 and GA2 are added to analyze the influence of the numerical resolution on the evolution of the magnetic field.
Furthermore, a special run of GA1 is performed to study the influence of the modeled star formation on the evolution of the magnetic field.
Finally, to study the influence of magnetic fields on the existing simulations, additional runs with only hydrodynamics are obtained.


\section{Simulations}

This section presents the results obtained from the numerical simulations introduced above.
Contour images of different quantities are created by projecting the \textsc{Spmhd} data in a comoving $(1$ $\rmn{Mpc})^{3}$ cube on a $512^{2}$ grid using the code \textsc{P-Smac2} (Donnert et al., in preparation).


\subsection{Morphological and magnetic evolution}

\begin{figure*}
\begin{center}
  \includegraphics[bb= 180 320 970 760, height=7cm,width=0.80\textwidth]{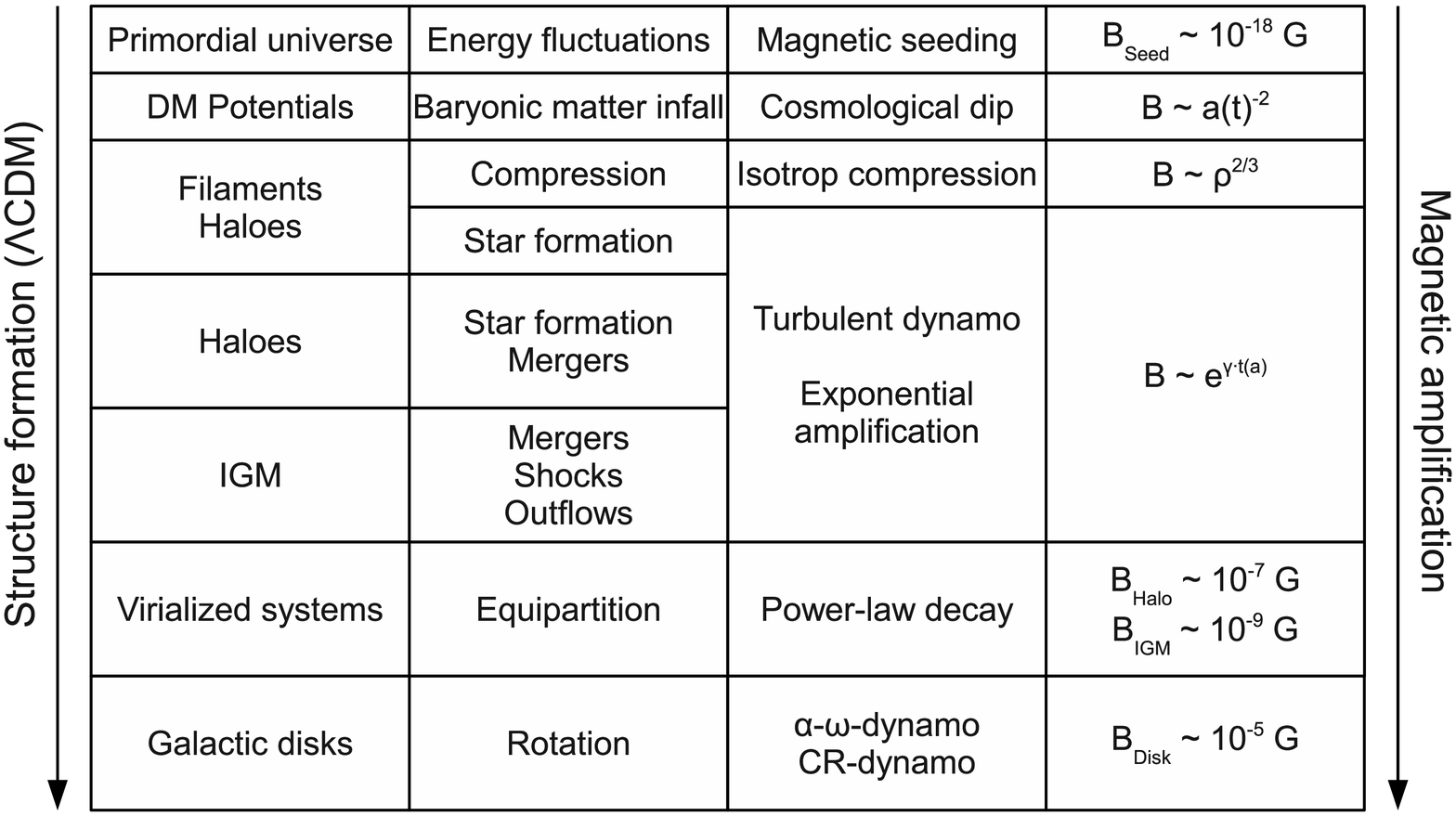}
  \caption{Overview of the different physical processes during structure formation.
This table shows the stages of the evolution (column 1), during which an astrophysical process (column 2) is triggering a MHD mechanism (column 3) operating on the magnetic field, resulting in the equations and magnetic field strength values given in column 4.}
  \label{fig:structure}
\end{center}
\end{figure*}

\begin{figure*}
\begin{center}
  \includegraphics[width=\textwidth]{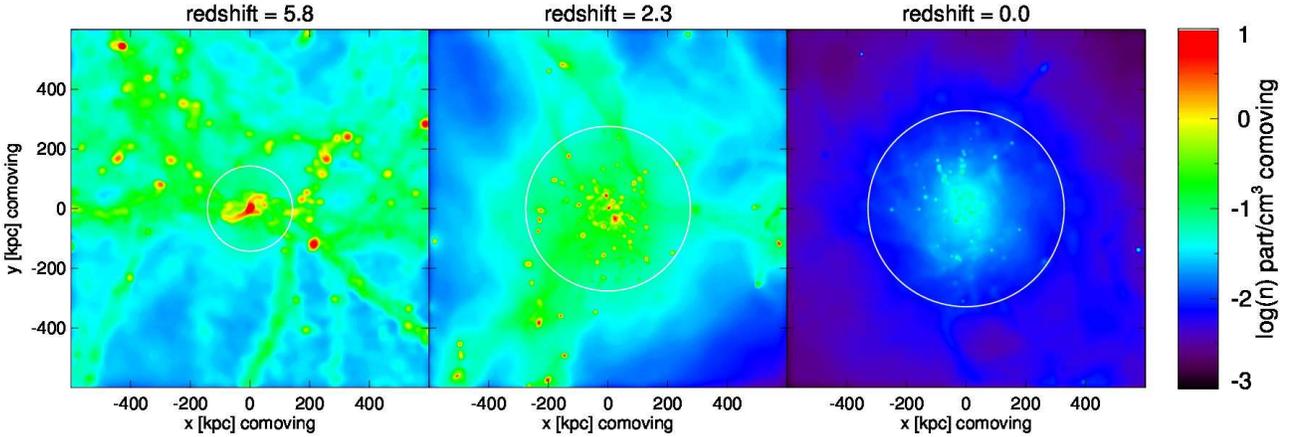}
  \caption{Projected number density $n_{\rmn{gas}}$ in comoving units at different redshifts in the simulation ga2\_bx18.
The shown regions are cubes with 1 Mpc (comoving) edge length centered on the halo center of mass.
The white circles indicate the virial radius of the halo.
The formation of filaments and protohaloes with subsequent merger events can be seen.}
  \label{fig:sim_dens}
\end{center}
\end{figure*}

\begin{figure*}
\begin{center}
  \includegraphics[width=\textwidth]{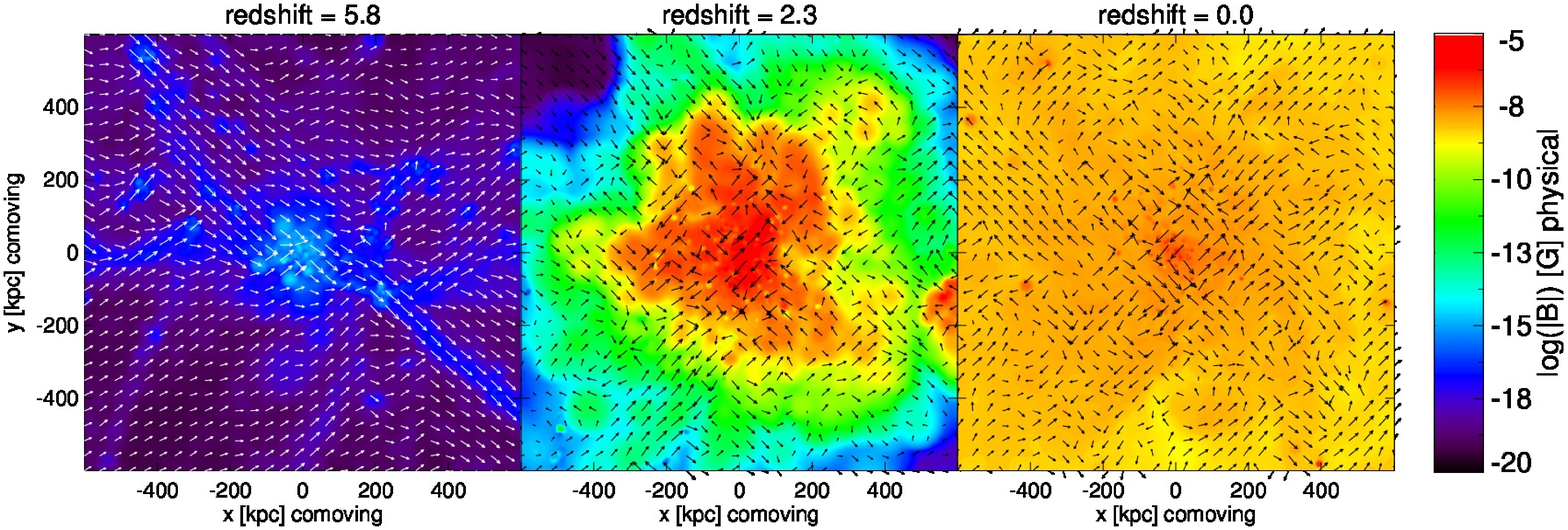}
  \caption{Projected total magnetic field strength and magnetic field vectors in physical units at different redshifts in the simulation ga2\_bx18.
The shown regions are cubes with 1 Mpc (comoving) edge length centered on the halo center of mass.
Clumping of the magnetic field together with the gas in filaments and amplification within protohaloes can be seen.
Furthermore, shockwaves are driven into the IGM increasing the magnetic field strength, until it saturates on all scales.}
  \label{fig:sim_bfld}
\end{center}
\end{figure*}

An overview of the different stages of structure formation and their implications on the magnetic field is shown in Fig. \ref{fig:structure}.
First, dark matter protohaloes and filaments are formed at redshift $z\approx{}30-10$, in which potential wells baryonic matter falls in.
Within these structures the frozen-in magnetic field gets compressed.
For isotropic compression this leads to $B\sim\rho^{2/3}$.
Note, that perpendicular compression of the magnetic field lines would lead to $B\sim\rho$.
Due to the cosmological expansion, the magnetic field strength outside these protohaloes decreases with $a^{-2}$.
Turbulence is mainly created by the gravitational collapse.
Secondly, as the gas density increases in the first structures, the threshold density is reached and star formation sets in at redshift $\approx$ 10, further enhancing the existing turbulence and consuming the available gas.
In these dense regions, small-scale dynamo action starts, increasing the magnetic field strength exponentially, i.e. $B\sim{}e^{\gamma{}t}$ with the growth rate $\gamma$ (see also section \ref{sec:analytical}).
Merger events will create shockwaves propagating into the IGM, possibly amplifying the magnetic field by compression and small-scale dynamo action.
As equipartition is reached, the system relaxes and turbulent motions will decay, with additionally the magnetic field decaying.

\noindent{}Fig. \ref{fig:sim_dens} shows the projected number density of the gas at three different redshifts in the simulation ga2\_bx18, together with the corresponding virial radii of the forming galactic halo.
Fig. \ref{fig:sim_bfld} shows the corresponding projected total magnetic field strength, as well as arrows indicating the direction of the magnetic field.
The different phases during the formation of the halo and the magnetic field amplification can clearly be seen.
The magnetic field agglomerates together with the gas in filaments and protohaloes, where small-scale dynamo action is taking place.
Furthermore, as merger events take place, shockwaves are propagating into the IGM creating turbulence.
The IGM magnetic field is amplified in stages with several shockwaves propagating into it.
Within each shockwave, the magnetic field is possibly amplified by compression within the shockfront and by small-scale dynamo action behind the shockfront \citep[see][for an analysis of shockwaves and their effect on the magnetic field during merger events]{kotarba11}.
At redshift $\approx$ 1 the last major merger event takes place and the magnetic field saturates, i.e. it evolves into energy density equipartition.
The magnetic field saturates inside the halo at a mean value of $\approx{}10^{-7}$ G and within the IGM at $\approx{}10^{-9}$ G, both at redshift 0.

\begin{figure}
\begin{center}
  \includegraphics[angle=90,width=0.475\textwidth]{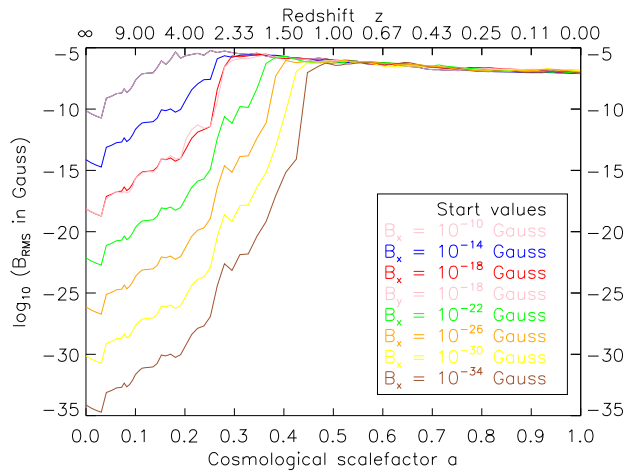}
  \caption{Growth curves of the volume-weighted RMS magnetic field strength inside the halo in the simulations GA0 for different seed fields.}
  \label{fig:sim_babs}
\end{center}
\end{figure}

\noindent{}Fig. \ref{fig:sim_babs} shows the RMS magnetic field strength inside the galactic halo as a function of redshift for different seed field strengths.
The amplification timescale (i.e. the gradient of the $\bmath{B}(a)$ function during the exponential amplification phase) is the same for all seed fields and only the total time until saturation varies.
After saturation, the magnetic field strength decreases with a power-law slope of $\approx{}-1$, as irregularities in the magnetic field are dissipated.

\begin{figure}
\begin{center}
  \includegraphics[angle=90,width=0.475\textwidth]{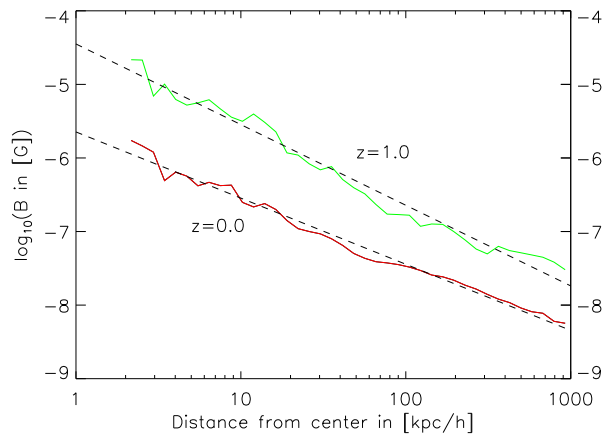}
  \caption{Radial profiles of the mean magnetic field strength inside the halo in the simulation ga2\_bx18 for redshifts 1 and 0, respectively.
For $z=1$, which is just after the last major merger event, the field strength decreases with a slope of -1.1.
For the relaxated system at $z=0$ the slope is -0.9.}
  \label{fig:sim_prof}
\end{center}
\end{figure}

\noindent{}Fig. \ref{fig:sim_prof} shows radial profiles of the volume-weighted magnetic field strength inside the galactic halo for two different redshifts.
At redshift 1 (after the last major merger), magnetic field strengths of several $10^{-5}$ G are reached in the center of the halo and drop to $\approx{}10^{-7}$ G at the virial radius with a slope of $\approx{}-1.1$.
At redshift 0 (virialized system with decaying turbulence), magnetic field strengths of several $10^{-6}$ G are reached in the center of the halo and drop to $\approx{}10^{-9}$ at the virial radius with a slope of $\approx{}-0.9$.
Since the gas density scales linearly with the distance from the galactic halo center, this indicates a relation of the form $B\sim\rho$.

\noindent{}Summing up, within the presented simulations it is possible to amplify a weak primordial magnetic field up to the observed equipartition values.

\subsection{Pressures and star formation}

\begin{figure*}
\begin{center}
  \includegraphics[bb= 160 340 730 830, height=9cm,width=0.80\textwidth]{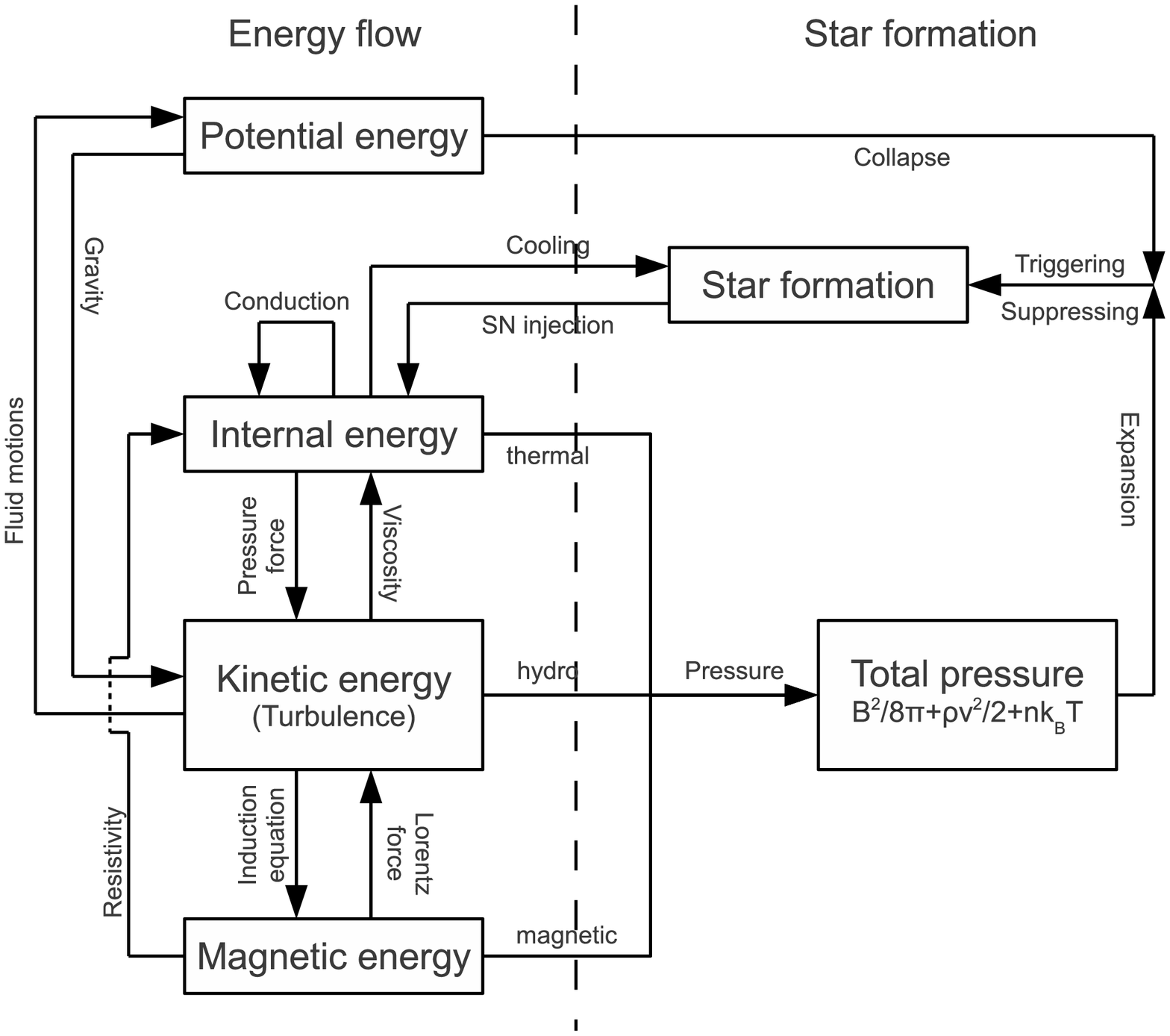}
  \caption{Diagram visualizing the flow of energy in the simulated MHD system.
The star formation model provides a source (SN injection) and a sink (cooling) of energy.
On the other hand, the different pressure components (thermal-, hydrodynamic and magnetic pressure) have an effect on the star formation.}
  \label{fig:energy_flow}
\end{center}
\end{figure*}

\noindent{}Fig. \ref{fig:energy_flow} summarizes the energy flow and its effect on the star formation within the simulations.
Via internal energy, star formation provides a sink (cooling) and source (SN injection) of internal energy of the gas.
Internal and kinetic energy are mutually exchanging via pressure forces and viscosity.
Additionally, the gravitational collapse transforms potential energy into kinetic energy, which is converted partly back into potential energy through the fluid motions in the potential wells.
Furthermore, kinetic motions create magnetic energy (induction equation).
The Lorentz-force describes the back-reaction of the magnetic field on the velocity field.
Non-ideal resistivity redistributes the magnetic energy and also converts it into internal energy.
Internal, kinetic and magnetic energy densities contribute to a total pressure.
The balance between the gravitational collapse and the total pressure support regulates the star formation rate.

\begin{figure}
\begin{center}
  \includegraphics[angle=90,width=0.475\textwidth]{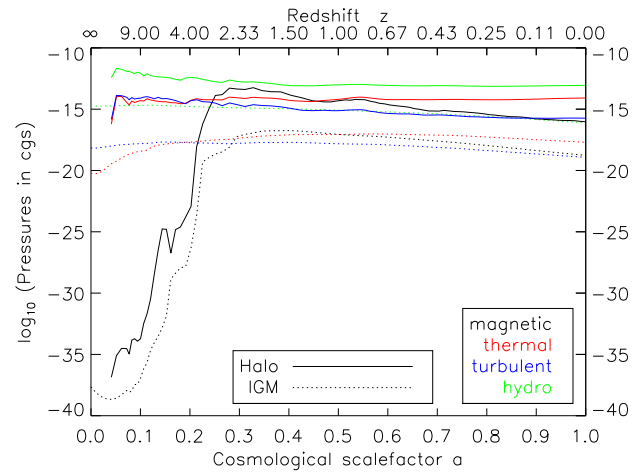}
  \caption{Volume-weighted energy densities as a function of redshift in the simulation ga2\_bx18 inside the halo and within the IGM.
The magnetic energy density (black line) gets amplified during the phase of halo formation until it reaches equipartition with the other energy densities, particularly the turbulent energy density (blue line).}
  \label{fig:sim_equi}
\end{center}
\end{figure}

\noindent{}Fig. \ref{fig:sim_equi} shows the magnetic energy density $\varepsilon_{\rmn{mag}}=B^{2}/8\pi$, the kinetic energy density $\varepsilon_{\rmn{kin}}=\rho{}v^{2}/2$, the thermal energy density $\varepsilon_{\rmn{therm}}=(\gamma-1)\rho{}u$ and the turbulent energy density $\varepsilon_{\rmn{turb}}=\rho{}v_{\rmn{turb}}^{2}/2$ in the simulation ga2\_bx18.
The adiabatic index $\gamma$ is $5/3$ and $u$ denotes the internal energy.
Similar as in \cite{kotarba10}, $v_{\rmn{turb}}$ is taken as an estimate of the turbulent velocity within the volume defined by a SPMHD particle.
They find it to be a good \textsc{Spmhd} approximation of the turbulent velocity, although it overestimates the turbulence on small scales and ignores the turbulence on scales larger than the smoothing scale.

\noindent{}As shown in Fig. \ref{fig:sim_equi}, the magnetic energy density increases from the seed value up to equipartition with the turbulent and thermal energy densities until a redshift of $\approx$ 3.
The cosmological dip can be clearly seen at the start of the simulations.
The magnetic energy density overshoots the turbulent energy density slightly in the beginning, which results from possible further compression after equipartition is reached.
Afterwards, the virialized system relaxes, and the turbulent and magnetic energies decline.
The thermal energy density still rises, as the magnetic and turbulent energy are converted into thermal energy by resistivity and viscosity.
With some delay, equipartition is also reached in the IGM by a stepwise amplification process:
First, the magnetic field is amplified inside the most dense structures, and subsequently the IGM magnetic field undergoes merger-driven shock amplification.

\begin{figure}
\begin{center}
  \includegraphics[angle=90,width=0.475\textwidth]{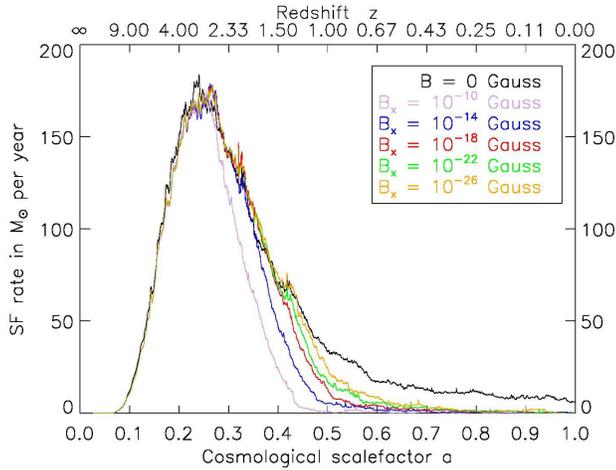}
  \caption{Total star formation rate as a function of redshift in the simulations GA0 with different magnetic seed field.
For simulations with magnetic fields, the star formation rate decreases when equipartition is reached and the additional magnetic pressure prevents the gas from reaching the density threshold required for star formation.}
  \label{fig:sim_sfr}
\end{center}
\end{figure}

\noindent{}Fig. \ref{fig:sim_sfr} shows the total star formation rate as a function of the cosmological scalefactor for all GA0 simulations (see Table \ref{tab:setup_sims}) with different seed field strengths.
Before the formation of filaments and protohaloes, no star formation takes place.
At the time the first structures reach the necessary critical density ($z\approx{}10$), star formation begins.
As more gas is accreted onto the main halo through gravitational infall or due to merger events, the star formation rate rises.
At a redshift of $\approx$ 3 it peaks and then starts to decline.
For the simulations without any magnetic fields, star formation is still constantly ongoing at a low rate by the end of the simulations.
For the simulations including magnetic fields, the star formation rate decreases earlier than in the non-magnetized comparison runs, as soon as equipartition is reached ($z\approx{}3)$, and is thus comperatively lower than in the comparison run or even stops completely by the end of the simulations.
The magnetic configuration at equipartition provides additional support against further gas accretion, preventing the gas inside the halo from reaching the threshold density necessary to form stars.

\noindent{}Note, that the details of star formation depend on the resolution of the simulation.
\cite{springel03b} show that when increasing the resolution of the simulation, the time when the first stars form is pushed towards higher redshifts.
Nevertheless, at low redshifts, the star formation rate converges to a value independent of the resolution.
Furthermore, the total formed stellar mass does not change with resolution.
These details in the star formation model also influence the turbulent dynamo action.
The starting point of the dynamo action depends strongly on the time when the first gas has collapsed, cooled and reached small enough spatial scales.
As soon as star formation sets in, additional supernova energy is injected into the system, leading to more turbulence and higher magnetic field growth rates.
The saturation value of the magnetic field strength depends on the turbulent energy density, which in turn depends on the the total injected feedback, and hence the total formed stellar mass.

\noindent{}The simulation ga1\_bx18\_nosf (not shown) is performed with the same setup and methods, but with disabled star formation module.
Within this simulation, the magnetic energy density does not get amplified up to equipartition with the turbulent energy density, but only rises a few orders of magnitude.
This is clearly, because radiative cooling lowers the internal energy of the gas, thus allowing the gas to clump more heavily and reach higher densities.
This results in smaller \textsc{Spmhd} smoothing lengths, and hence the small-scale dynamo will also operate on smaller scales, thus leading to higher growth rates.
Summing up, radiative cooling and supernova feedback are important in MHD simulations of galactic halo formation.

\subsection{Numerical reliability}\label{sec:stability}

\begin{figure}
\begin{center}
  \includegraphics[angle=90,width=0.475\textwidth]{plots/divb_err.eps}
  \caption{Mean numerical magnetic divergence measure $\left<h\left|\bmath{\nabla}\cdot\bmath{B}\right|/|\bmath{B}|\right>$ in ga2\_bx18 inside the halo and within the IGM.}
  \label{fig:num_div}
\end{center}
\end{figure}

\noindent{}The numerical magnetic divergence $\left<h\left|\bmath{\nabla}\cdot\bmath{B}\right|/|\bmath{B}|\right>$ is a common measure regarding the reliability of \textsc{Spmhd} simulations \citep[e.g.][]{price12}.
It is calculated for every particle $i$ inside its kernel, which is a sphere with radius equal to the smoothing length $h_{i}$.
Even for simulations employing the Euler potentials, which are free of physical divergence by construction, this measure can reach values of the order of unity \citep{kotarba09}.
The numerical divergence can be regarded as a measure for quality of the numerical calculations and the irregularity of the magnetic field inside each kernel and is not related to possible physical divergence \citep{kotarba10,buerzle11a}.
Here, an estimator for the numerical divergence of the form

\begin{equation}\mathrm{NumDivB}_{i}=\sum_{j}{\frac{h_{i}+h_{j}}{|\bmath{B}_{i}|+|\bmath{B}_{j}|}\frac{m_{j}}{\rho_{i}}\left(\bmath{B}_{i}-\bmath{B}_{j}\right)\cdot\nabla{}W(r_{ij},h_{i})}\end{equation}

\noindent{}is used, where $W$ is the \textsc{Spmhd} kernel function between the particle $i$ and its neighbours $j$.

\noindent{}Fig. \ref{fig:num_div} shows the mean numerical divergence within the halo (solid line) and within the IGM (dashed line) as a function of scalefactor for the simulation ga2\_bx18.
Throughout the entire simulation, the numerical divergence remains below unity.
The numerical divergence is zero in the beginning of the simulations, as expected for a uniform magnetic seed field.
During the phases of merger events and magnetic field amplification, the error estimator rises.
The turbulent dynamo tangles the magnetic field lines and creates irregularities also below smoothing scales, resulting in a non-vanishing numerical divergence.
During the phase of relaxation of the halo, NumDivB decreases, as the field lines are unfolded and ordered and the magnetic energy is dissipated or transferred to larger scales.
Within the IGM, the numerical divergence remains constant.
This is because within the IGM the NumDivB-decreasing process of magnetic field reordering is balanced by NumDivB-increasing processes.
These are the accretion of gas onto the halo and the expansion of space, both resulting in an increasing smoothing length and thus an increasing NumDivB.


\section{Agreement of model and simulations}\label{agreement}

\noindent{}Since the simulations start at a finite redshift and not at $z=\infty$, a simple shift of $B^{2}_{0}=a^{-4}_{\rmn{start}}B^{2}_{\rmn{start}}$ for the initial magnetic field strength in Eq. (\ref{equ:solv_final}) is used.
The characteristic turbulent quantities in Eq. (\ref{equ:kulsrud_gamma}) are assumed to be constant in space and time.
This approximation is good for the phase of strong star formation between redshift $10$ and redshift $1$, where also the majority of the magnetic amplification is taking place, while later the growth rate is truncated, and the form of the turbulent quantities is negligible.

\begin{figure}
\begin{center}
  \includegraphics[angle=90,width=0.475\textwidth]{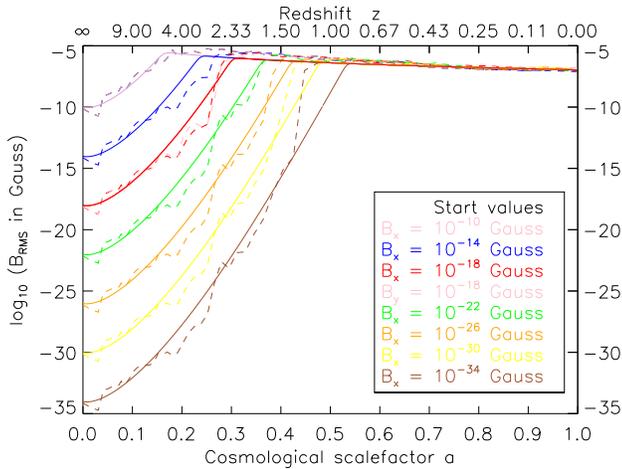}
  \caption{Analytical growth functions (solid lines) as given by Eq. (\ref{equ:solv_final}) using the parameters listed in Table \ref{tab:model_parameter} for different magnetic seed field strengths.
After an initial cosmological dip, the magnetic field is growing exponentially until it reaches equipartition with the turbulent energy density.
Additionally, the simulated growth cruves from Fig. \ref{fig:sim_babs} are shown (dashed lines).}
  \label{fig:model_plot}
\end{center}
\end{figure}

\noindent{}In Fig. \ref{fig:model_plot} simulated growth curves of the magnetic field strength are shown (dashed lines).
These curves match notably well with the calculated curves (solid lines).
Table \ref{tab:model_parameter} shows the numerical values used for the calculations, which are resulting in a timescale ($e$-folding time) of $\approx{}90$ Myr for the growth of the large scale magnetic field.
Extracting such values directly from the simulations is quite challenging, as the density, velocity and length scales within the simulated galactic haloes range over many orders of magnitude.
However, the length scale $l_\rmn{turb}$ on which the magnetic energy density increases first within the main halo is determined by the size of star forming regions, which (within the main halo) can be associated to substructures.
Such substructures (clumps of gas, stars and dark matter) can be identified using \textsc{Subfind} \citep{springel01b,dolag09a}.
Only the largest substructures still contain gas and form stars within the main halo, having masses of $\approx{}(10^{8}-10^{10}) M_{\odot}$ and thereby diameters of a few $10$ kpc.
They are orbiting with typical velocities of a few 100 km s$^{-1}$, inducing gas RMS velocities within the main halo between 50 and 100 km s$^{-1}$ during the time of rapid growth of the halo, where also the magnetic amplification is taking place (between $z\approx10$ and $z\approx1$).
Such values strongly motivate the choice of $l_{\rmn{turb}}\approx25$ kpc and $v_{\rmn{turb}}\approx75$ km s$^{-1}$ leading to the good agreement between the simulations and the analytical model.

\begin{figure}
\begin{center}
  \includegraphics[angle=90,width=0.475\textwidth]{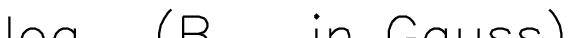}
  \caption{Volume-weighted RMS magnetic field strength inside the halo as a function of redshift for three simulations with the same magnetic seed field of $10^{-18}$ G but different resolution, together with the analytical growth function (see Eq. (\ref{equ:solv_final}) and Table \ref{tab:model_parameter}).
The analytical function fits the simulated evolution of the halo magnetic field very well.}
  \label{fig:model_compare}
\end{center}
\end{figure}

\noindent{}Fig. \ref{fig:model_compare} shows a comparison of the analytical growth function (red line) as calculated according to Eq. \ref{equ:solv_final} using the values given in Table \ref{tab:model_parameter} together with the halo RMS magnetic field strength in the simulations ga0\_bx18, ga1\_bx18 and ga2\_bx18, respectively.
All curves start with the same magnetic seed field of $B_{\rmn{start}}=10^{-18}$ G.
The resolution of the simulations increases by each a factor of roughly 10 from ga0 to ga1 and from ga1 to ga2, respectively.
Compared to the simulation with the standard resolution (brown line), the analytical growth function fits the simulated magnetic field evolution very well.
Also, for higher resolutions (black and green lines), the fit is convincing.
Note, that for higher resolutions, star formation sets in at higher redshifts leading to an earlier rise of collapsed, cooled gas and feedback and hence an earlier starting point for the turbulent dynamo.
Nevertheless, the saturation value of the magnetic field strength is indistinguishable.

\noindent{}For the calculation of the hydrodynamic and magnetic Reynolds numbers, the typical velocity $V$ and length scale $L$ are determined by the physical properties of the system, i.e. the soundspeed $c_{\rmn{s}}^{2}=\gamma{}(\gamma{}-1)u$ and the Alfv\'{e}nspeed $v_{\rmn{a}}^{2}=\bmath{B}^{2}/\mu_{0}\rho$, and the size of the halo, respectively, which do not depend on resolution.
Within all the simulations, a constant turbulent resistivity $\eta_{\rmn{turb}}$ is used, and hence, $Rm$ stays constant.
Additionally, artificial viscosity as given by \cite{price12} is applied, where $\nu$ depends on particle properties and spacings.
A higher resolution thus leads to smaller $\nu$, and thus (given the constant $V$ and $L$) to higher $Re$ numbers and hence more turbulence.
A higher (or, better resolved) turbulence in turn results in a higher growth rate of the magnetic field.

\noindent{}The cosmological turbulent dynamo as described by Eq. (\ref{equ:solv_final}) reproduces the main features of the simulated non-ideal magnetic field amplification very well.


\section{Summary}

\noindent{}In this paper the evolution of magnetic fields during galactic halo formation is discussed. 
The main focus is placed on the investigation of the processes responsible for the amplification of magnetic fields from seed field levels to observed values.
An analytical model for the evolution of the magnetic field driven by a turbulent dynamo is presented and the predictions of this model are compared with numerical, cosmological simulations of Milky-Way like galactic halo formation including the evolution of magnetic fields, radiative cooling and star formation.
The most important results are summarized as follows:

\begin{itemize}
\item A primordial magnetic seed field of low strength can be amplified up to equipartition with other energy densities during the formation and virialization of a galactic halo in a $\Lambda$CDM universe.
The final magnetic field strength decreases with a slope of $\approx$ -1.0 from $\approx{}10^{-6}$ G in the center to $\approx{}10^{-9}$ G behind the virial radius (IGM) of the halo and also reaches $\approx{}10^{-5}$ G in interacting systems.
These values are in notably good agreement with observations \citep{beck96,kronberg08}.

\item The magnetic field amplification in filaments and protohaloes is dominated by turbulent dynamo action.
Radiative cooling of the primordial gas is needed in order to reach spatial scales small enough for the turbulent dynamo to operate efficiently.
Equipartition is reached on small scales first and later on larger scales, consistent with theoretical expectations \citep{brandenburg05,shukurov07,arshakian09}.
The turbulence is driven by the gravitational collaps, by supernova activity and by mergers of protohaloes into the main galactic halo.
After equipartition is reached, the magnetic energy decays with a power-law dependance of $B_{t} \sim t^{-4/3}$.
The IGM magnetic field is amplified by outflows of magnetized gas from the center of the haloes and by merger-driven shock amplification outside the main halo.

\item The amplification timescale ($e$-folding time) of the order of $10^{7}$ yr is small enough to describe the generation of strong magnetic fields in irregular galaxies at high redshifts as observed \citep[e.g.][]{bernet08}.

\item The structure of the resulting magnetic field is random and turbulent.
Additional dynamo processes, e.g. the $\alpha$-$\omega$ dynamo \citep{ruzmaikin79,shukurov07} or the cosmic-ray-driven dynamo \citep{lesch03,hanasz09} are needed to produce regularity in the magnetic field topology.

\item Last, but not least, a basic analytical model is able to reconstruct the numerical results very accurately.
Weak magnetic perturbations grow in a non-stationary turbulent hydrodynamical flow.
This amplification is slightly modified by the $\Lambda$CDM cosmology, particularly in the early universe.
Non-ideal truncation of the growth rate finally yields equipartition.
These processes result in an analytical growth function which fits the simulations astonishingly well.
\end{itemize}

\noindent{}In the current picture of galaxy formation -- together with cooling and star formation -- magnetic fields are efficiently amplified from seed field levels up to the observed values.
However, their detailed influence on the dynamics of the gas and the underlying seeding mechanisms still remain unclear and need to be investigated further.


\section*{Acknowledgments}
We thank the anonymous referee for the comments, which helped to improve many parts of this paper.
Special thanks to Felix Stoehr for providing the original initial conditions.
Rendered graphics are created with P-Smac2 (Donnert et al., in preparation).
F.S. is supported by the DFG Research Unit 1254.
K.D. is supported by the DFG Priority Programme 1177 and by the DFG Cluster of Excellence 'Origin and Structure of the Universe'.


\bsp

\label{lastpage}

\end{document}